# Aerosol Characteristics at a High Altitude Location in Central Himalayas: Optical Properties and Radiative Forcing

**P. Pant, P. Hegde, U. C. Dumka, Ram Sagar**

Aryabhatta Research Institute of Observational Sciences, Manora Peak, Nainital, India.
(ppant@aries.ernet.in)

**S. K. Satheesh**

Centre for Atmospheric and Oceanic Sciences, Indian Institute of Science, Bangalore, India
(satheesh@caos.iisc.ernet.in)

**K. Krishna Moorthy**

Space Physics Laboratory, Vikram Sarabhai Space Centre
Thiruvananthapuram, India.
(krishnamoorthy_k@vssc.org)

*Abstract.* Collocated measurements of the mass concentrations of aerosol black carbon (BC) and composite aerosols near the surface were carried out along with spectral aerosol optical depths (AODs) from a high altitude station, Manora Peak in Central Himalayas, during a comprehensive aerosol field campaign in December 2004. Despite being a pristine location in the Shivalik Ranges of Central Himalayas, and having a monthly mean AOD (at 500 nm) of 0.059 ±0.033 (typical to this site), total suspended particulate (TSP) concentration was in the range 15 - 40 μg m$^{-3}$ (mean value 27.1 ± 8.3 μg m$^{-3}$). Interestingly, aerosol BC had a mean concentration of 1.36 ± 0.99 μg m$^{-3}$, contributed to ~5.0 ± 1.3% to the composite aerosol mass. This large abundance of BC is found to have linkages to the human activities in the adjoining valley and to the boundary layer dynamics. Consequently, the inferred single scattering albedo lies in the range of 0.87 to 0.94 (mean value 0.90 ± 0.03), indicating significant aerosol absorption. The estimated aerosol radiative forcing was as low as –4.2 W m$^{-2}$ at the surface,

+0.7 W m$^{-2}$ at the top of the atmosphere, implying an atmospheric forcing of +4.9 W m$^{-2}$. Though absolute value of the atmospheric forcing is quite small, which arises primarily from the very low AOD (or the column abundance of aerosols), the forcing efficiency (forcing per unit optical depth) was ~88 W m$^{-2}$, which is attributed to the high BC mass fraction.

## 1. Introduction

The importance of radiative forcing of the atmosphere by aerosols is assuming increasing importance in climate studies [*Haywood and Ramaswamy*, 1998; *Jacobson*, 2001; *Ramanathan et al.*, 2001; *Kaufman et al.*, 2002]. Nevertheless, large uncertainties still persist in this area, primarily due to the large heterogeneity in aerosol properties and scarcity of adequate data. In this context the Asian region assumes importance due to its high population density and diverse human activities. Numerous measurements and impact assessments are being reported from this region in the recent years [*Moorthy et al.*, 1999, 2004; *Satheesh and Ramanathan*, 2000; *Ramanathan et al.,* 2001; *Babu et al*., 2002; *Seinfeld et al.*, 2004; *Pandithurai et al*., 2004; *Tripathi et al*., 2005]. However, most of these studies focused to either urban/semi-urban landmass or oceans adjacent to densely populated coastal belt. Investigations from remote, sparsely inhabited regions have the importance of providing a sort of background against which the urban impacts can be compared. In this perspective, observations from high altitude stations have a special significance as the aerosols in this region provide a 'far-field picture', quite away from potential sources and are more representative of free troposphere conditions. By free troposphere, we mean region largely unaffected by the meso-scale planetary boundary layer dynamics. The properties of such aerosols will be representative of a greater spatial extent too. Recent observations have revealed pockets of high aerosol loading / optical depth in the north Indian regions around the Ganga basin, particularly during the winter season [*Girolamo et al.*, 2005; *Tripathi et al.*, 2005; *Jethva et al*., 2005] when the prevailing meteorological conditions are favorable for confinement of aerosols. The topography of the region with the towering Himalayas to the north adds to this. The low-level capping inversion, particularly during winter periods of highly reduced solar insolation, prevents the dispersion of pollutants.

In this paper, we report the results of collocated measurements of spectral aerosol optical depths (AOD), mass and number concentrations of composite aerosols, and mass concentration of aerosol black carbon (BC) from a high altitude station in the central Himalayas during a field campaign, conducted from $1^{st}$ to $30^{th}$ December 2004. The measurements are used to characterize the aerosols and to assess their radiative impacts. The implications are discussed.

## 2. Observational Details and Data

### 2.1. Site Description and Synoptic Meteorology

The campaign was carried out from Manora Peak ($29.4^o$ N; $79.5^o$ E, 1950 m AMSL), Nainital, located in the Shivalik ranges of Central Himalayas (marked as 1 in Fig. 1). Owing to its large elevation this site is mostly above the boundary layer particularly during winter season (December to February), when the temperature drops very low (as low as 5° C) and the thermal convection is weak. As such, this region is well above the convective boundary layer, the height of which is typically <~800 m during this season [ eg., *Moorthy et al.,* 2004; *Tripathi et al.*, 2005]. . The prevailing meteorology during December comprises of a synoptic westerly circulation with a weak northerly component (see Fig. 2), low ambient temperature, and low to moderate relative humidity (RH), and scanty rainfall. The mean wind speed, temperature and RH characteristics during the campaign period, as measured at the site using an automatic weather station, are shown in Fig. 3. The winds are quite weak (2 to 4 m s$^{-1}$); mean temperature varies between 8 to 15° C and RH varies between 50 to 70%. The sky was generally clear and cloud-free except for three days. No rainfall occurred during the campaign period. Even though occurrence of fog conditions were reported in the later half of December in the north Indian plains, lying 200 kms due south (of Manora Peak), and persisted for several days, no fog occurred at the observational site on any of the campaign days. The topography of the region is shown in Fig. 4 and vegetation index in Fig. 5. The sharply peaking topography of the Himalayan ranges is clearly seen from Fig. 4. Locations at an aerial distance of < 40 kms due south and south west of the location are at very low elevation (<200 m AMSL) and merge with the vast plain level of the Ganga basin, while to the north and north east we have the tall Himalayan ranges. The plains are densely populated and include several urban areas like New Delhi, Kanpur (marked as 2 in Fig.1), Lucknow within aerial distance of 200 kms. However, the observation site is well below the snow line and has thick vegetation as can be seen from Fig. 5.

**2.2. Experimental details and Data**

During the experimental campaign simultaneous and collocated measurements of spectral AODs ($\tau_{a\lambda}$), BC mass concentration ($M_B$), mass loading of composite aerosols (TSP) near-surface ($M_T$), number concentration of composite aerosols in the size range from 0.3 to 20 μm, and meteorological parameters were carried out respectively using a 5-channel Microtops Sun photometer (Solar Light Co., USA), Aethalometer (Magee Scientific, USA model AE-21), a high volume air sampler (Envirotech Inc., model APM 430), an optical particle counter (OPC, model 1.018 of Grimm Aerosol Technik, GmbH, Germany), and an automatic weather station (Campbell Scientific Inc., Canada). The Aethalometer uses a continuous filtration and optical transmission technique to yield $M_B$ in near real time. It aspirates ambient air (from 5 meter above ground level) using its inlet tube and pump. The BC mass concentration is estimated by measuring the change in the transmittance of its quartz filter tape, on to which the particles impinge. More details on the instrument and principle of measurement are available elsewhere [*Hansen et al.,* 1984*, Hansen,* 2003]. The Aethalometer was operated at a standard flow rate of 3.3 litres min$^{-1}$ and a time base of 5 min. As the standard flow rate correspond to a temperature of 293 K and pressure of 1017 hPa, the instrument measured values of $M_B$ were corrected for the changes in the pump speed at the high altitude (mean pressure of 850 hPa and temperature of 12°C during the observation period) following the considerations given in *Moorthy et. al.* [2004]. The mass loading ($M_T$) of total suspended particulates (TSP) was measured using the high volume air sampler. Since, the aerosol concentrations at the site were very low, as for most of the cases, the HVS was operated about 16 hours to obtain one sample.

The OPC is a portable, lightweight, battery operated optical particle counter, which classifies particles into 15 size channels in the diameter range between 0.3 to 20 μm (Model No 1.108 of Grimm Aerosol Technik, GmbH, Germany; http://www.grimm-aerosol.com) and estimates the number concentration. It works on the principle of optical counting of particles, by detecting the pulses of radiation scattered (at 90°) by individual particles in the sampling flow. A diode laser is used as the light source. The pulse height discriminator sets the size range and the number of pulses corresponds to the number of particles. The OPC has been operated in the count mode and provided number concentration ($N_C$). An averaging time of 5 min was used to obtain the number concentration. An isokinetic probe connected to the instrument inlet aspirated the ambient air at a rate of 1.2 litre per minute. The high dynamic range of the instrument enables measurements from clean to polluted environments. This instrument was mounted in its environmental housing, which maintained the inlet air sample at a constant RH level.

Aerosol optical depth (AOD) were estimated using Microtops II Sunphotometer which is a hand-held instrument having 5-channels with central wavelengths at 380, 440, 500, 675 and 870 nm and FWHM (full width at half maximum) of ~6 - 10 nm. The methodology of data acquisition with Microtops II and precautions during measurements are described in detail by *Porter et al.* [2001], *Morys et al.* [2001] and *Ichoku et al.* [2002].

## 3. Results and Discussion

### 3.1. Aerosol Optical Depth

Spectral aerosol optical depth ($\tau_{a\lambda}$) measurements were made regularly during periods of unobscured solar conditions and more than 650 data sets were obtained during the campaign period. These are averaged in daily ensembles to obtain the daily mean. The temporal variation in daily mean AOD at 500 nm is shown in Fig. 6(a) where the vertical bars represent the standard deviations. The optical depths were very low, with $\tau_{a\lambda}$ at 500 nm being less than 0.1 on all days, except on December 01, 18 and 21; thereby indicating the prevalence of the clean environment. These AODs are considerably lower than the values (0.2 to 0.34) reported for winter 2003, by *Ramana et al*. [2004], based on observations from four high altitude sites in Nepal, due southeast of Manora Peak in the Himalayan ranges. This is typical to this location during the winter season (December to February) as has been reported earlier [*Sagar et al.*, 2004]. The monthly-mean AOD at 500 nm was 0.059 ±0.033. The mean AOD spectrum for December 2004 is shown in Fig. 6(b). By evolving a least square fit to the individual AOD spectra (on log-log scale) and the Angstrom parameters ($\alpha$ and $\beta$) are evaluated. The values of $\alpha$ varied between 0.4 to 1.2 on different days while $\beta$ varied from 0.02 to 0.05. The mean value of $\alpha$ for December 2004 was 0.09. The regional distribution of the monthly mean AOD (at 550 nm) derived from MODIS (onboard Terra satellite) is shown in Fig. 7. The picture clearly shows low AODs over the Himalayan regions and regions of enhanced AODs (by an order of magnitude) on the adjoining valley. This spatial variation is not a simple artifact of the altitude alone. If we assume an exponential variation with altitude for the aerosol extinction with a typical scale height of 2 km [*Chami and Santer,* 1998*; Levy et al.,* 2004], then the observed mean value of 0.059 for AOD at 500nm at ~2km altitude of the Manora peak will translate to a value of ~0.16 for sea level. Rather, if we consider a much lower scale height of ~1.5 km for the aerosols for the winter season (in view of the largely reduced convection), the AOD still would be only 0.22 for the sea-level and this value is considerably lower than those reported for several continental sites in India (remote and semi-urban) for this season, where the AODs typically are ~0.3 to 0.5 at 500nm. The relevance of this spatial heterogeneity in AOD has implications to the aerosol properties over Manora peak as will be seen in the following sections.

### 3.2. Concentrations of Composite and Black Carbon aerosols

Daily average values of BC concentration show considerable variation ranging from 0.85 to 2.6 µg m$^{-3}$. In Fig. 8(a) we show the monthly mean diurnal variation of $M_B$ while the day-to-day variations are given in Fig. 8(b). The monthly mean BC concentration was found to be 1.36 ±0.99 µg m$^{-3}$, which is very low in comparison to other urban locations in India (Fig. 1) as reported by *Babu et al.* [2002] for Bangalore (BLR); *Babu and Moorthy* [2002] for Trivandrum (TVM); *Lata and Badrinath* [2003] for Hyderabad (HYD); and *Tripathi et al.* [2005] for Kanpur (KNP). This value of $M_B$, though less than the values reported at these locations in India (TVM, HYD, BLR, KNP, marked in Fig.1); is still quite high when we consider the elevated nature of the station and its remoteness. Part of this BC could be transported from the valley below, by the boundary layer dynamics, while the rest would be the prevailing values and those resulting from the local activities, which though is limited. The monthly mean diurnal variation at Manora Peak shows lower values of $M_B$ during night and morning hours and increases slowly thereafter approaching its highest level late in the afternoon (around sun set) and then decreases more rapidly to reach the low value around midnight. Diurnal minimum occurs at ~7:00 am. The amplitude of the diurnal variation is ~ 3; from ~0.75 to 2.5 µg m$^{-3}$ and is comparable to that seen at other continental locations [eg., *Babu and Moorthy*, 2002]. However, the nature of the diurnal variation observed at the low altitude locations were different from (or in fact opposite to) that observed at Manora peak. For example, *Babu and Moorthy* [2002] have reported high BC values at night and morning hours, which gradually decreased as the day advances, attaining its minimum level around 16:00 local time and recovered into the night. On the contrary, at Manora Peak we see a late afternoon high and a midnight minimum. It may be recalled that based on BC measurements from a high altitude station (1250 m AMSL), *Bhugwant et al* [2001] have reported diurnal variations of BC very similar to that seen in this study and is attributed to the boundary layer dynamics which brings up the polluted emissions from the underlying valley regions to this high-altitude location.

To the northeast of the Manora Peak is the hilly terrain of the Central Himalayan ranges (>2.0 km AMSL), having insignificant anthropogenic activities. On the other hand to the south and southwest are densely populated valley regions (at a mean elevation of ~300 m), and located at an aerial distance of ~10 km. During nighttime, the shallow nocturnal boundary layer acts as a capping inversion [*Stull,* 1999], confining the emissions from these regions well below the mountain peak and as a result the BC concentration at Manora peak, is mainly due to the residual layer (of the preceding day). After the sunrise, the heating of the land surface results in setting up of convective motions, which gradually raise the inversion to higher altitudes [*Vladimir and Eischinger,* 2004], leading to increased vertical mixing that brings up the air with high aerosol concentration to higher levels. This would result in the increase in the concentration of anthropogenic emissions at the peak as is observed in the present study. It would be so, not only for the BC concentration, but also for the number/ mass concentrations of the composite aerosols. This is examined in Fig. 9, where we have plotted the average diurnal variation of the number concentration of the composite aerosols (recorded by the OPC), for December 2004. The striking similarity with Fig 8 (a) is apparent. These observations indicate that the vertical transport of aerosols from the densely populated valley regions (which also have high AODs as seen in Fig. 7) by the evolving boundary layer, significantly modifies the concentration of aerosols at the peak. As a result, the afternoon (AN) concentrations of aerosol species (when the boundary layer has fully evolved) become higher compared to the forenoon (FN) concentrations (when the boundary layer is shallow). To examine this, we have separated $M_B$ and $N_C$ into forenoon and afternoon (FN and AN) parts of the day and obtained separate averages for each day. The day-to-day variations of these, shown in Fig. 10, very clearly indicate consistently higher concentrations in the afternoon. The mean values (separately for FN and AN) for the month are listed in Table 1 for $M_B$ and $N_C$.

As far as the aerosol absorption is concerned, more than the mass concentration ($M_B$) of BC, it is the mass mixing ratio ($F_{BC}$) (which is the ratio of the mass concentration of BC to that of the composite aerosol = $M_B/M_T$) is important. Atmospheric forcing efficiency of aerosols is shown to be strongly dependent on $F_{BC}$ [*Babu et al.*, 2004]. From the simultaneous measurements of the mass concentrations of composite and BC aerosols (respectively $M_T$ and $M_B$) using the HVS and Aethalometer respectively, the BC mass fraction $F_{BC}$ was estimated for each day. $M_T$ was found to be in the range of 15 to 40 µg m$^{-3}$, with a mean value of 27.1 ±8.3 µg m$^{-3}$. The average share of BC to composite aerosol mass concentration at Manora peak was found to be ~ 5.0 ± 1.3%. It is much lower than the values reported for Kanpur as 7-15% [*Tripathi et al.,* 2005] and for Trivandrum ~12% [*Babu and Moorthy,* 2002] during the month of December. The evaluated $F_{BC}$ at Manora Peak is lower than even the lowest values reported by various investigators during the INDOEX program [*Satheesh et al.,* 1999; *Babu et al.,* 2002; *Venkataraman et al.,* 2002].

Investigations during ACE-2, INDOEX and TARFOX [*Novakov et al.,* 2000; *Ramanathan et al.,* 2001; *Satheesh et al.,* 2002] have shown BC to contribute in the range of 10 to 15% to the composite aerosol mass, whereas its mixing ratio reported over tropical western Pacific was as high as 25% (with BC mass of ~40 µg m$^{-3}$) [*Liley et al.,* 2002]. During the field campaign at an urban site, Bangalore (13°N, 77°E, 960 m msl) in India, large concentration of BC was observed; both in absolute terms (~6 µg m$^3$) and fraction to composite aerosol mass (~11%). From an extensive and continuous measurement of aerosol

BC from a coastal location, Trivandrum, adjacent to the Arabian Sea, *Babu and Moorthy* [2002] observed that BC mass fraction to the total showed large seasonal variations. During winter (similar to the present study period) BC contributes to ~11% of the aerosol mass (4 to 6 µg m$^{-3}$ in absolute terms) while during monsoon season the BC fraction is reduced to 3 to 4% (1 to 2 µg m$^{-3}$ in absolute terms). Measurements during 1996-97 close to down town Helsinki, have shown BC concentration in the range of 1.2 to 1.5 µg m$^{-3}$ with an average of 1.3 µg m$^{-3}$ and its average contribution to aerosol mass was 19% [*Pakkanen et al.,* 2000]. Examining in the light of the above the average concentration of BC and its mass fraction to the composite aerosols at Manora Peak are lower than those reported at several locations for similar season; nevertheless the values are rather higher than what would be expected for such an elevated and remote location.

**3.3. Single Scattering Albedo and Aerosol Radiative Forcing**

The magnitude of the climate forcing by aerosol particles is strongly dependent upon the aerosol single scattering albedo (SSA), which is the ratio of the scattering to extinction coefficients. The aerosol characteristics estimated in earlier sections were used to estimate SSA and direct short wave radiative forcing. For this, we have followed an approach of using the available observations as anchoring points in a standard continental average aerosol model, which is fine-tuned to match the measurements [see *Moorthy et al*., 2005 for details]. The optical and radiative properties, estimated from this model were then fed to a radiative transfer model. This method has been extensively used for estimating radiative forcing due to aerosols [for eg., *Satheesh et al.,* 1999; *Podgorny et al*., 2000; *Conant et al*., 2003; *Markowicz et al*., 2003; *Moorthy et al*., 2005]**.** Based on observations of aerosol optical depth, Angstrom wavelength exponent, total aerosol mass and BC mass, we have adopted the continental average model of *Hess et al* [1998] and used the BC mass fraction, aerosol optical depths and Angstrom exponent as the anchoring points. The other components of the model were sulphate and transported dust. The measured BC was used to represent soot. The number density of each of these species was adjusted, while maintaining the mass fraction of BC to the observed value so that the spectral AODs of the modelled aerosols (assuming spherical aerosols and are assumed externally mixed) agreed well with the mean values from the measurements and the value of α agreed within 5%. Because of this closure (with spectral AOD) and anchoring the BC mass fraction, the initial assumption of the model will not have any significant impact on the estimated forcing, as has been pointed out by *Satheesh and Srinivasan* [2005]. The spectral AOD thus estimated from the aerosol model is shown by dotted line in Fig. 6b. The rms deviation between the experimentally estimated AOD spectrum and the model simulation is 0.006, indicating a good agreement between the two. The inferred SSA was in the range of 0.87 to 0.94 with a mean value of $0.9 \pm 0.03$. This value is significantly higher than that reported by *Babu et al.* [2002] for Bangalore (~0.78) and *Tripathi et al.* [2005] for Kanpur (~0.76). From sun-sky radiometer measurements, *Pandithurai et al.* [2004] have estimated the SSA over an urban site (Pune) to be ~0.81 for the same season of 2001-02. Based on near-surface measurements in the Himalayan region (in Nepal), *Ramana et al.* [2004] reported the SSA in the range of 0.7 to 0.9 for winter of 2004. Based on the observations over tropical Indian Ocean, *Satheesh et al* [1999] have reported a value of 0.9 during pre-INDOEX campaign in 1998. *Ramanathan et al.* [2001] reported a value of $0.86 \pm 0.08$ for the west coast of India and $0.89 \pm 0.02$ for Arabian Sea, and 0.95 to 0.99 for Southern Indian Ocean. The value of SSA at Manora Peak is comparable to the highest value reported for Himalayan region based on surface measurements and to that obtained over polluted oceanic regions, but is lower than that for the pristine environment (southern Indian Ocean).

The aerosol optical properties (single scattering albedo and phase function thus estimated following the above procedure) along with the measured spectral AODs and Angstrom wavelength exponent were used to estimate the short wave (0.25 to 4 µm), aerosol radiative forcing by incorporating in a Discrete Ordinate Radiative Transfer (RT) model developed by University of Santa Barbara (SBDART) [*Ricchiazzi et al*., 1998 for details]. The methodology used to estimate radiative forcing from measured aerosol properties is available extensively in the literature [*Satheesh et al*., 2002; *Moorthy et al*., 2005]. The estimated aerosol radiative forcing at the surface was as low as –4.2 W m$^{-2}$ and the corresponding atmospheric forcing was +4.9 W m$^{-2}$ (Fig. 11). The overall uncertainty in the estimated forcings, due to the small deviations in the simulations and uncertainties in the surface reflectance values used (from MODIS data), works out to be ~15 to 20%.

Clear-sky aerosol surface forcing reported was about -26 W m$^{-2}$ for Atlantic Ocean during TARFOX [*Hignett et al.,* 1999] and -29 W m$^{-2}$ for tropical Indian Ocean during INDOEX [*Podgorny et al.,* 2000*; Satheesh and Ramanathan,* 2000]. Using measurements during 1996 cruise of ORV Sagar Kanya and island experiments, *Jayaraman et al.* [1998] and *Conant* [2000] reported comparable values for tropical Indian Ocean. The TOA forcing over Indian Ocean was -10 W m$^{-2}$; comparable to the value of ~-9 W m$^{-2}$ over north Atlantic [*Satheesh and Ramanathan,* 2000*; Bergstrom and Russell,* 1999]. Using collocated data from the visible IR scanner (VIRS) and CERES (TRMM) satellite [*Christopher et al.,* 2000] made observational estimates of short wave TOA aerosol forcing as -68 W m$^{-2}$ in 1998 during a biomass-burning episode in Central America. During 1985 biomass burning over South America, regional aerosol radiative forcing values were estimated (using satellite data) and were in the range of -25 to -34 W m$^{-2}$ [*Christopher et al.,* 1998]. Aerosol radiative impact at the surface was studied using AERONET data in Brazil; Saharan dust in Cape Verde; and urban-industrial pollution in Creteil, near Paris, France, and near Washington, D. C [*Kaufman et al.,* 2002]. The measured 24-h average aerosol impact on the solar flux at the surface per unit optical depth was -80 W m$^{-2}$ in these sites, almost independent of the aerosol type: smoke, dust, or urban-industrial pollution [*Kaufman et al.,* 2002]. A field experiment was conducted to quantify aerosol direct radiative forcing in the southeastern United States (North Carolina) and at an adjacent valley site during June 1995 to mid-December 1995. The mean cloud-free 24-h aerosol forcing at the TOA (at the surface) for highly polluted, marine and continental air masses are estimated to be -8 ± 4 (-33 ± 16), -7 ± 4 (-13 ± 8), and -0.14 ± 0.05 (-8 ± 3) W m$^{-2}$, respectively. Based on investigations from Nepal *Ramana et al.* [2004] have observed a surface radiative forcing of –25 W m$^{-2}$ for the same season of 2003. Compared to the above, the aerosol radiative forcing at Manora Peak is significantly lower. Nevertheless, the forcing efficiency is quite large (~88 W m$^{-2}$) and is comparable to that over several polluted regions suggesting that type of aerosol over Manora Peak have a large efficiency for forcing, but the absolute forcing is low due to their low abundance in the column.

## 4. Conclusions

1. A comprehensive campaign to characterize the properties of aerosols was conducted at a high altitude region in the Central Himalayas during winter 2004. Columnar AODs were generally very low with a mean value of 0.059 ±0.033 at 500 nm, which is typical for this remote, high altitude location.

2. The estimated value of $\alpha$ ranged from 0.4 to 1.2 with a mean value of 0.9.

3. The concentrations of BC as well as composite aerosols showed significant diurnal variations, with an afternoon peak and low around mid-night; the amplitude of the diurnal variation was ~3. The afternoon peak is attributed to the lifting up of pollutants from the densely populated valley region, adjoining the peak, by the convective eddies. The monthly mean concentrations (of composite aerosols and BC) are higher by a factor of ~2 in the afternoon, as compared to the forenoon.

4. The inferred single scattering albedo was in the range of 0.87 to 0.94 with a mean value of 0.9 ± 0.03.

5. The estimated aerosol radiative forcing at the surface was as low as –4.2 W m$^{-2}$ and the corresponding atmospheric forcing was +4.9 W m$^{-2}$. Aerosol radiative forcing at Nainital was found to be much lower compared to those observed over low altitude sites in the Indo-Gangetic plain and southern India.

*Acknowledgements.* This work was carried out as a part of Land-Campaign-II, of the Indian Space Research Organisation's Geosphere Biosphere Program (ISRO-GBP). The authors would like to express their sincere gratitude to Prof. M.M Sarin, Physical Research Laboratory, Ahmedabad for providing the High Volume Sampler data for total mass concentration estimates.

**Table-1.** Changes in the monthly mean concentrations from FN to AN

| Period | $M_B$ (ng m$^{-3}$) | $N_C$ (m$^{-3}$) |
|---|---|---|
| Forenoon (FN) | 1140 ± 116 | (7.23 ± 0.78) x 10$^{10}$ |
| Afternoon (AN) | 2280 ± 204 | (1.66 ± 0.17) x 10$^{11}$ |

**Table – 2.** Comparison of SSA at Various Locations

| Location | Period | Nature of Location | SSA (500 nm) | Reference |
|---|---|---|---|---|
| Bangalore | Post Monsoon (November 2001) | Urban | 0.78 | Babu et al. [2002] |
| Kanpur | Winter (December 2004) | Industrialised Urban | 0.76 | Tripathi et al. [2005] |
| North Indian Ocean | Dec-February 1998 | Polluted Marine | 0.90 | Satheesh et al. [1999] |
| Coastal Arabian Sea | February-March 1999 | Polluted Marine | 0.86 | Ramanathan et al. [2001] |
| Central Arabian Sea | February-March 1999 | Polluted Marine | 0.89 | Ramanathan et al. [2001] |
| Southern Arabian Sea | February-March 1999) | Polluted Marine | 0.90 | Ramanathan et al. [2001] |
| Pune | Winter 2001-02 | Industrialised Urban | 0.81 | Pandithurai et al. [2004] |
| Manora Peak (present study) | Winter (December 2005) | High Altitude Clean | 0.90 | Present Study |

**Figure captions:**

Figure 1. Location of the study area along with other urban/ semi-urban stations in India, where BC concentration has been measured and reported in this paper (1: Manora Peak, 2: Kanpur, 3: Kharagpur, 4: Mumbai, 5: Hyderabad, 6: Bangalore, 7: Trivandrum).

Figure 2. Monthly mean wind vectors at 700 hPa level during December 2004, relevant to the observation site.

Figure 3. Monthly mean diurnal variations of wind speed (WS), temperature (Temp) and relative humidity (RH) for December 2004 with the standard error of each point.

Figure 4. Digital elevation model for the study region, showing the topography. The white circle shows the location of Manora Peak

Figure 5. Monthly mean NDVI distribution over India, with emphasis to the Central Himalayas for December 2004. (Data obtained from MODIS)

Figure 6. Day to day variations of daily-mean AOD at 500 nm based on Sun Photometer measurements (on the top panel) and the monthly mean AOD spectrum (in the bottom panel). The vertical bars are the standard deviations of the respective means.

Figure 7. MODIS-derived AOD (at 550 nm) for December 2004.

Figure 8. Monthly-mean diurnal variation of BC (top panel) and temporal variation of daily mean $M_B$. The error bars represent the standard deviation from the means.

Figure. 9. Monthly mean diurnal variations of the number concentration (per litre) of composite aerosols obtained using the optical particle counter.

Figure. 10. Day to day variations of the mass and number concentrations of BC and composite aerosols ($M_B$ and $N_C$) respectively. The points are the daily mean values and the vertical bars the standard deviation of the daily means. The monthly mean values are given by the horizontal lines (continuous for AN and dashed for FN part) and the respective standard errors are given by the error bars over them.

Figure 11. Comparison of aerosol radiative forcing observed over Manora Peak, Kanpur and Bangalore.

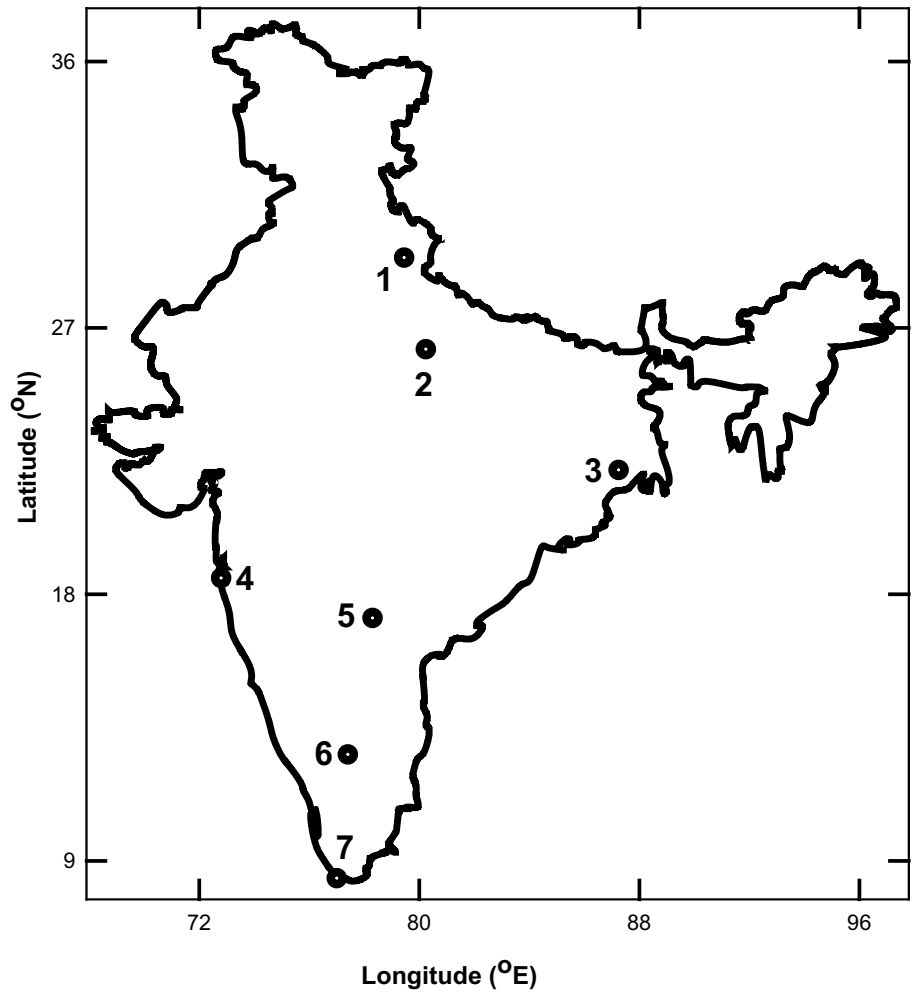

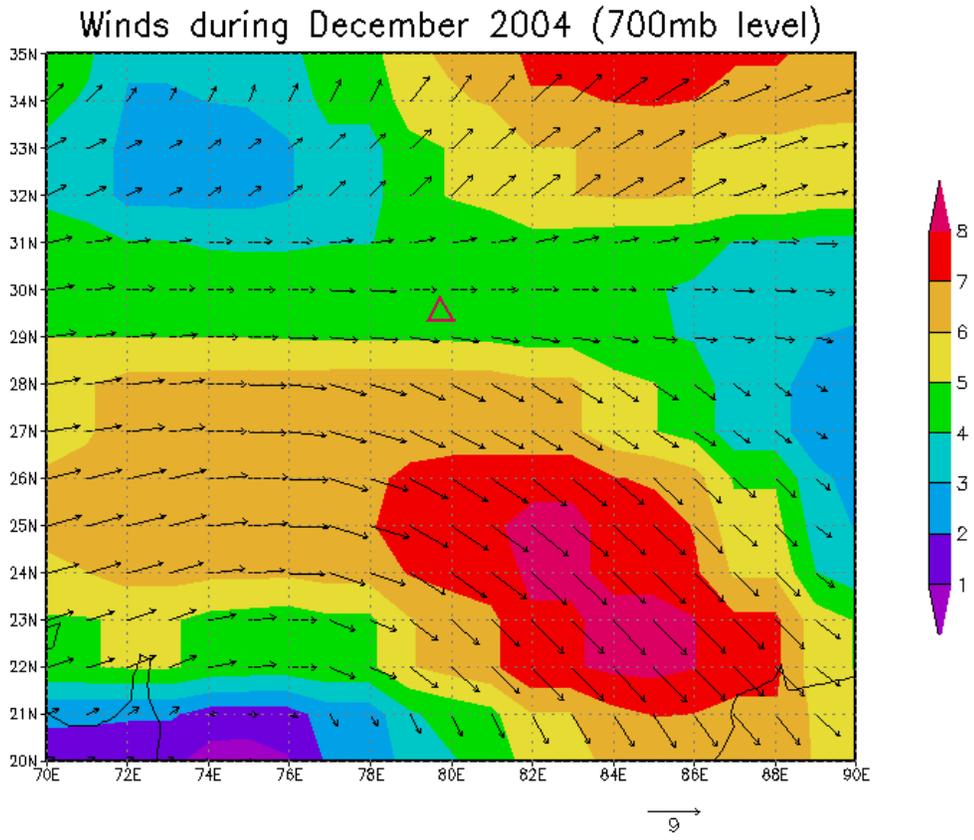

**Fig. 2.**

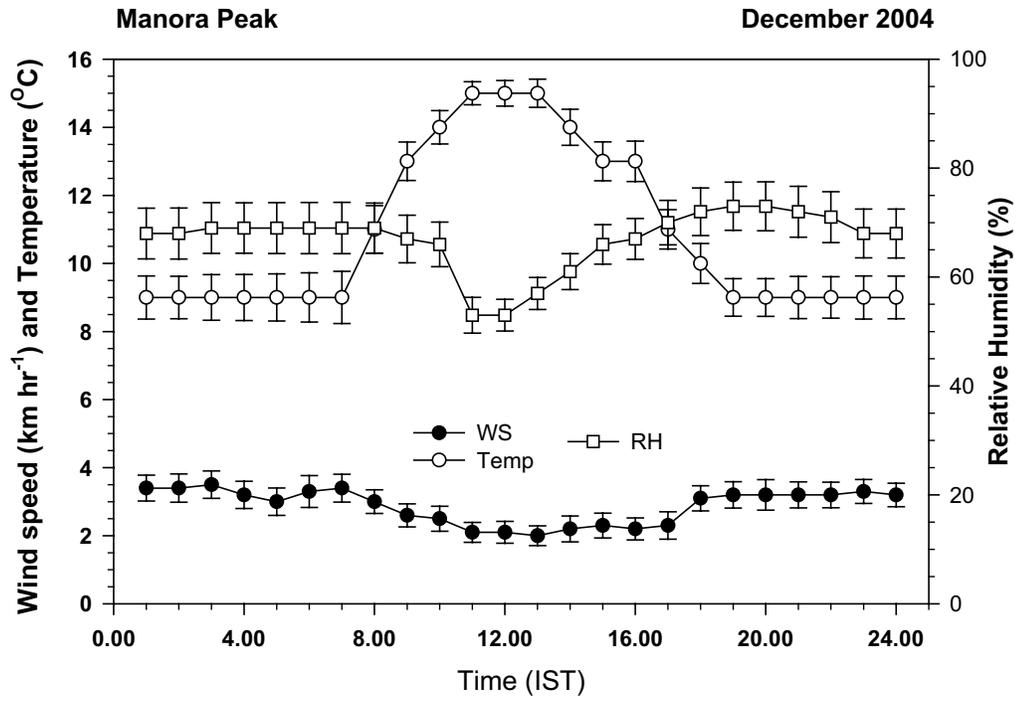

**Fig. 3.**

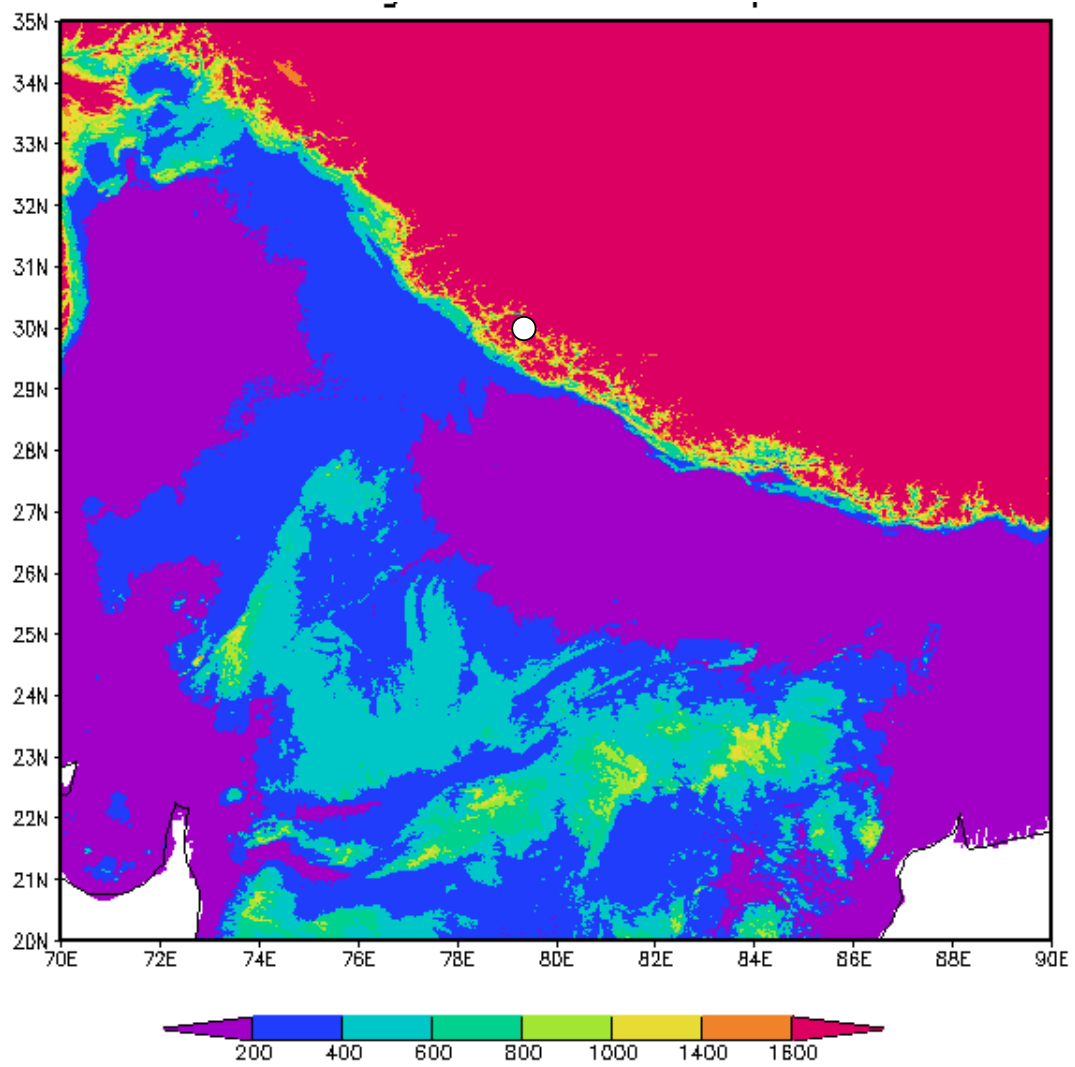

**Fig. 4.**

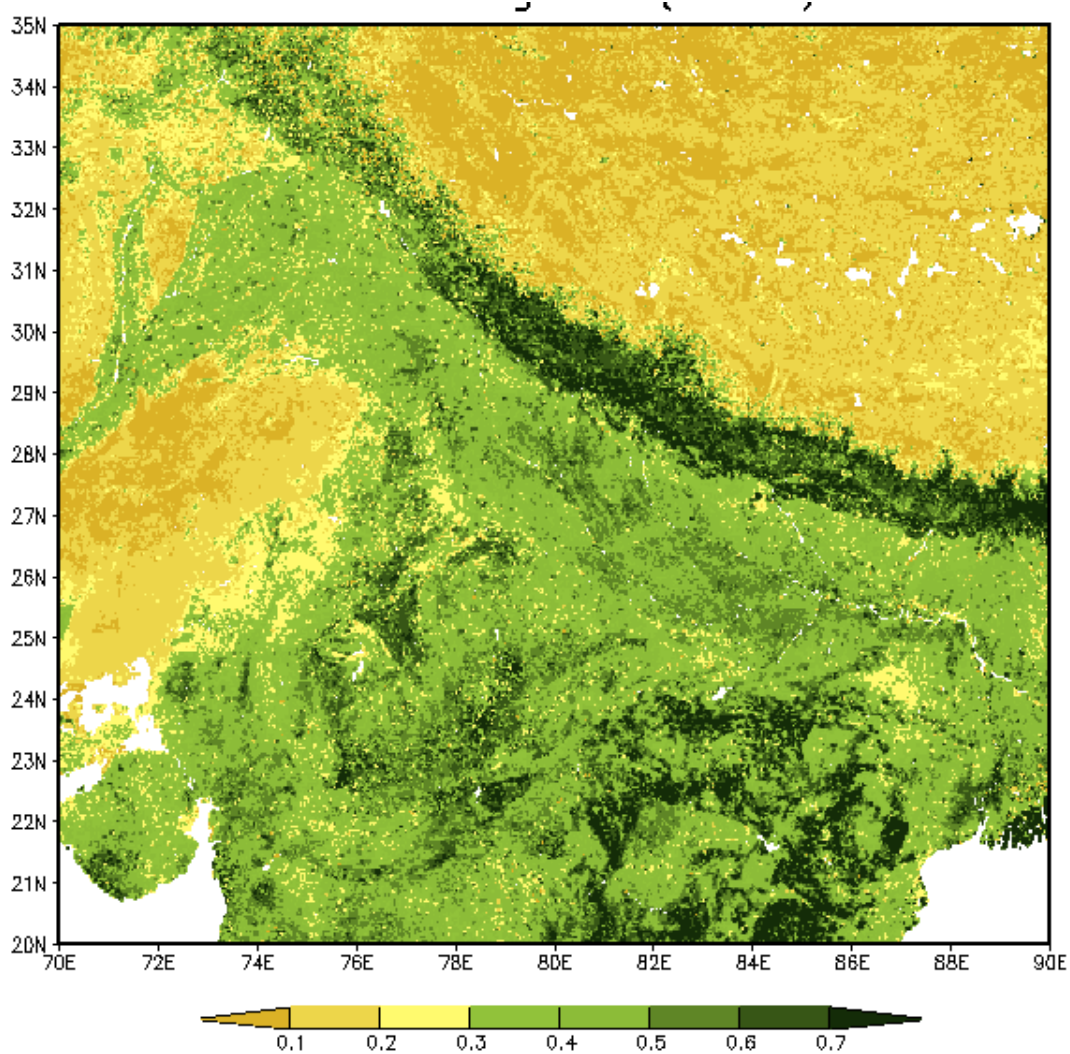

**Fig. 5.**

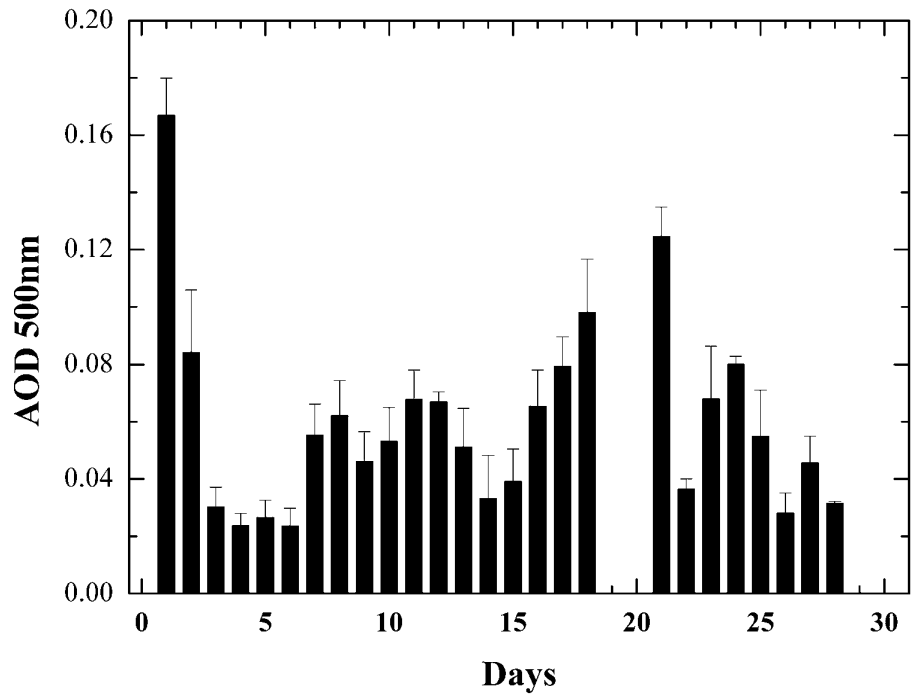
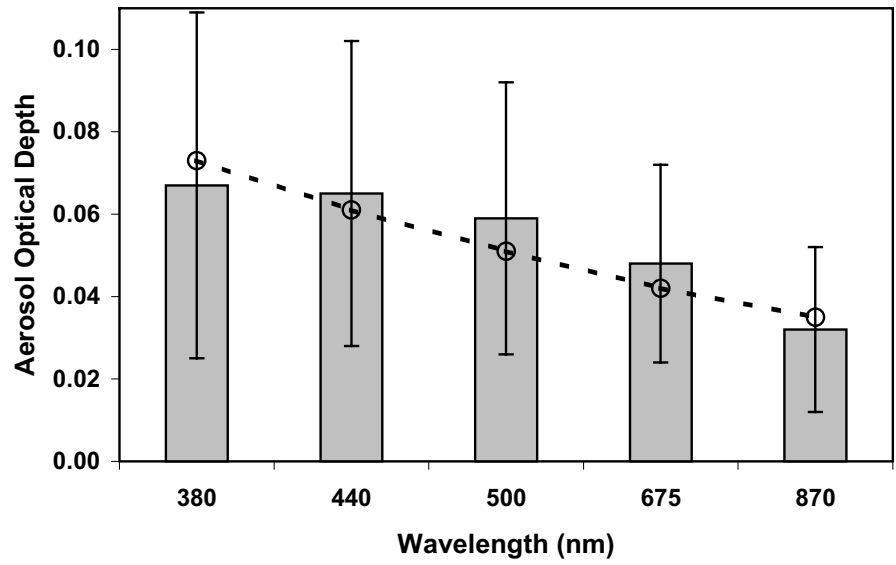

Fig. 6.

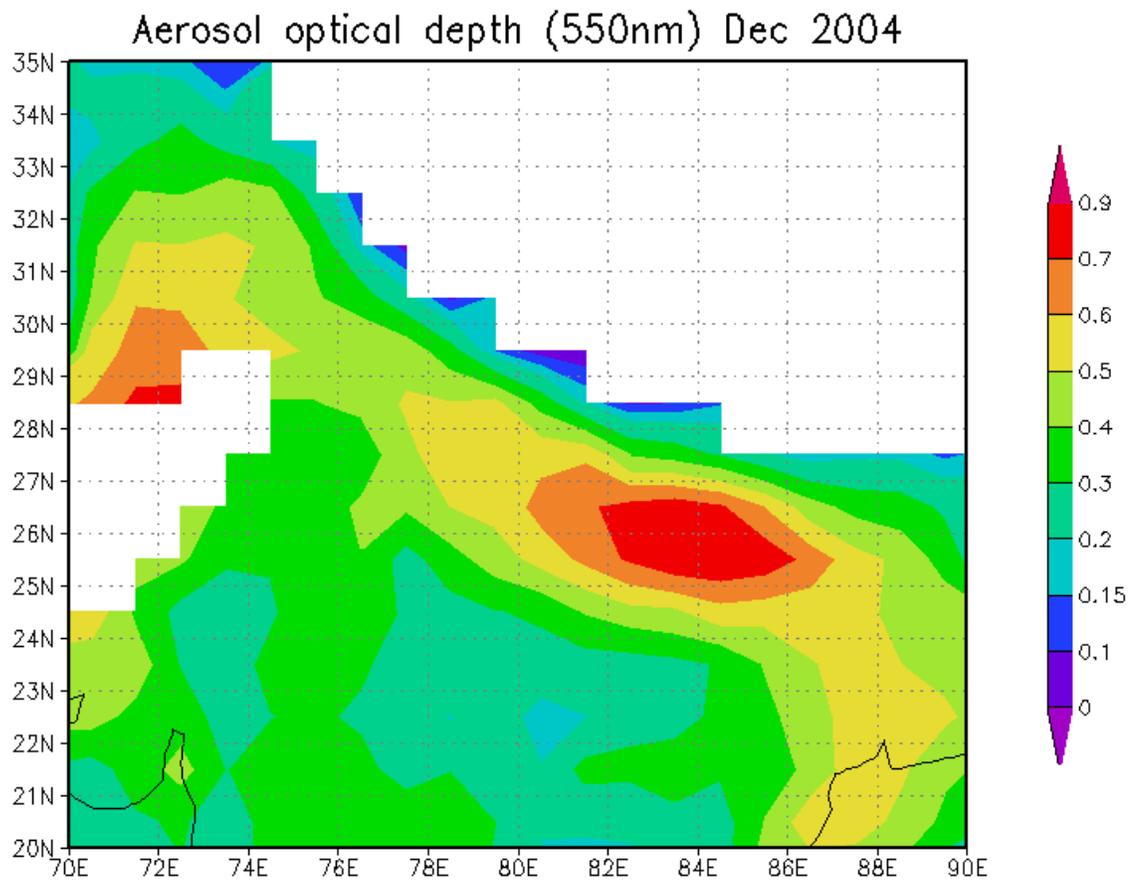

**Fig. 7.**

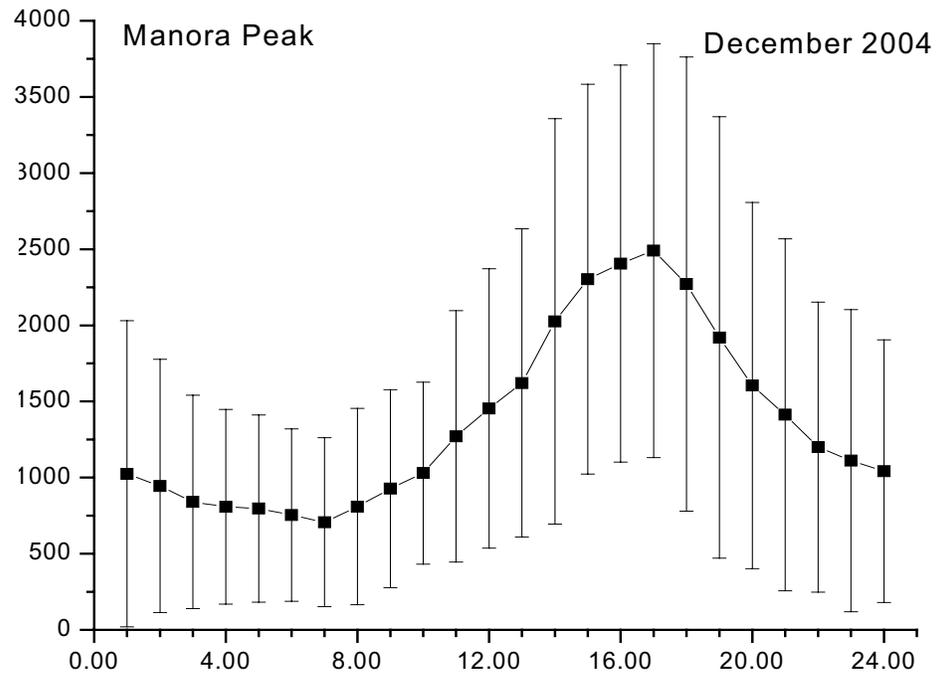

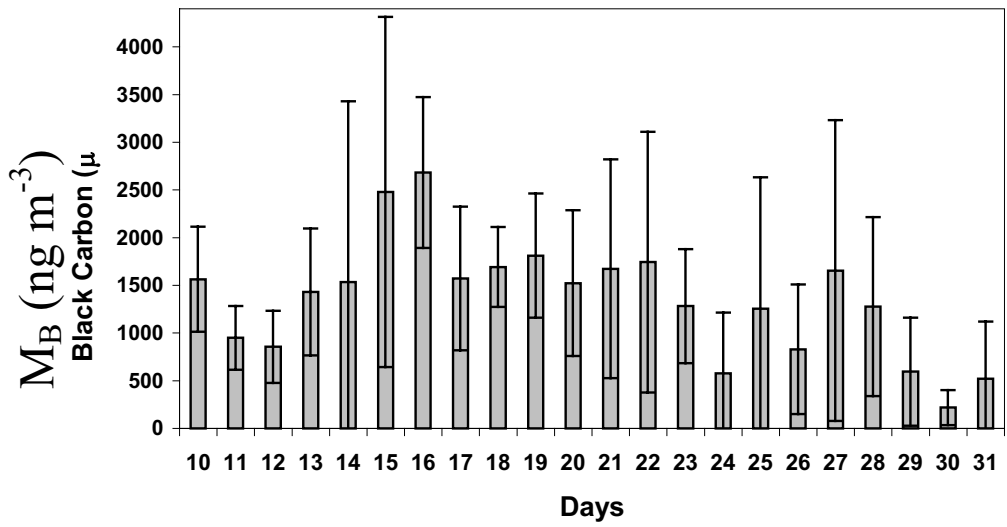

**Fig. 8.**

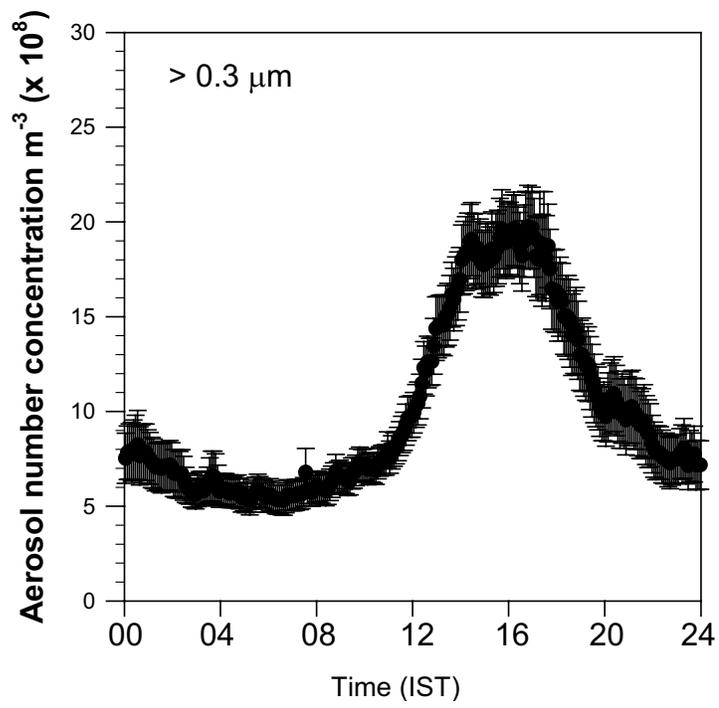

**Fig. 9**

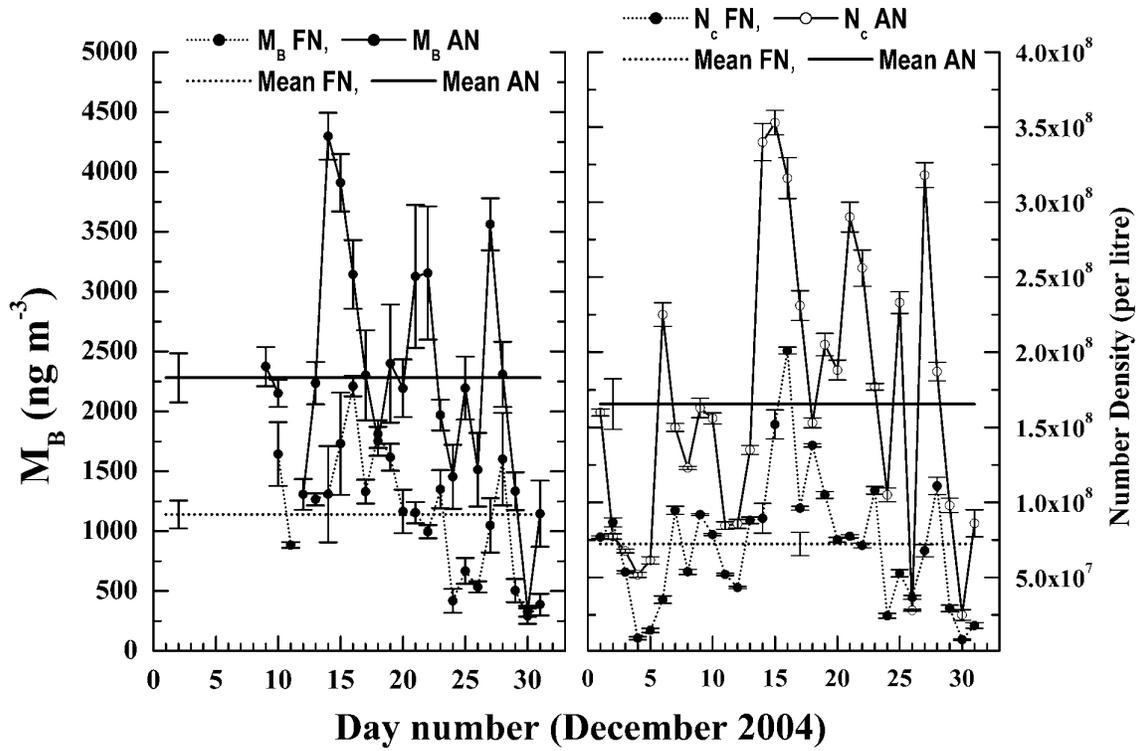

Fig. 10

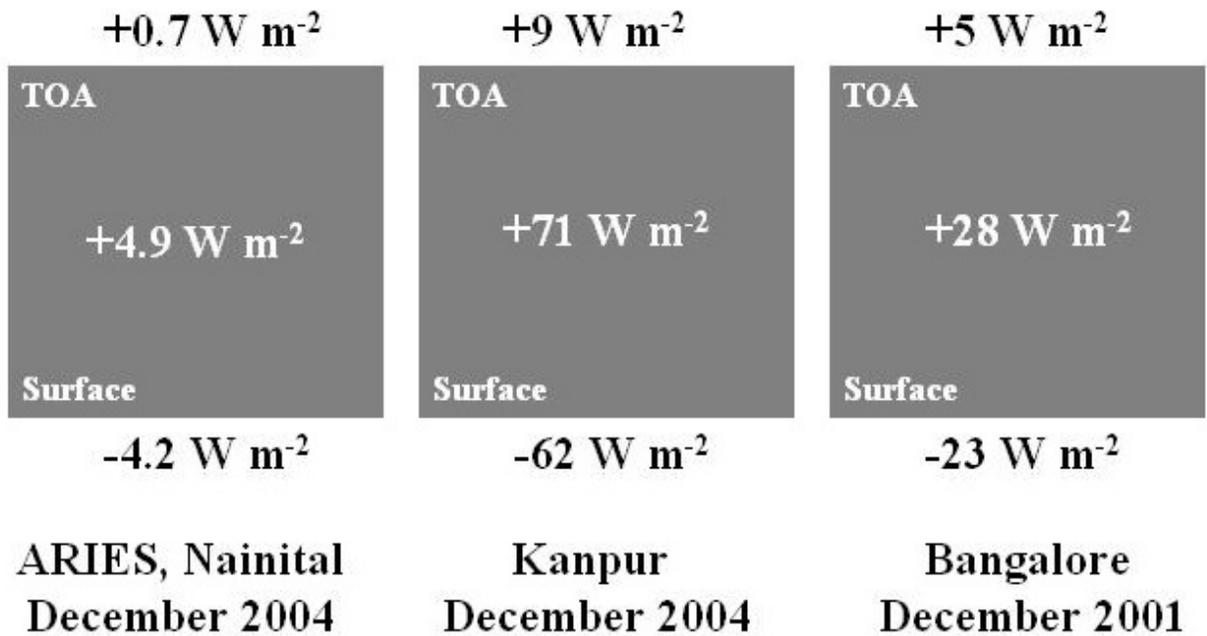

**Fig 11**